\documentclass[pre,aps,twocolumn,showpacs,floatfix,10pt,superscriptaddress]{revtex4-1}

\usepackage[english]{babel}
\usepackage[utf8]{inputenc}
\usepackage{amsmath}
\usepackage{graphicx}
\usepackage[colorinlistoftodos]{todonotes}
\usepackage{graphicx}
\usepackage{color}
\usepackage{amsmath, amsthm, amssymb}
\usepackage{dcolumn}
\usepackage{hyperref}
\usepackage{subfigure}
\usepackage{epsfig}
\usepackage{fancybox}
\newcommand{\beq}{\begin{equation}}
\newcommand{\eeq}{\end{equation}}	
\newcommand{\beqa}{\begin{eqnarray}}
\newcommand{\eeqa}{\end{eqnarray}}
\newcommand{\besm}{\begin{smallmatrix}}
\newcommand{\eesm}{\end{smallmatrix}}
\newcommand{\be}{\begin{equation}}
\newcommand{\ee}{\end{equation}}	
\newcommand{\bea}{\begin{eqnarray}}
\newcommand{\eea}{\end{eqnarray}}

\begin{document}

\title{A novel Recurrence-Transience transition and Tracy-Widom growth in\\ a cellular automaton with quenched noise} 
\author{Rahul Dandekar}
\email{dandekar@theory.tifr.res.in}
\affiliation{Institute of Mathematical Sciences - HBNI, Chennai, India}

\date{\today}

\begin{abstract}
We study the growing patterns formed by a deterministic cellular automaton, the rotor-router model, in the presence of quenched noise. By the detailed study of two cases, we show that: (a) the boundary of the pattern displays KPZ fluctuations with a Tracy-Widom distribution, (b) as one increases the amount of randomness, the rotor-router path undergoes a transition from a recurrent to a transient walk. This transition is analysed here for the first time, and it is shown that it falls in the 3D Anisotropic Directed Percolation universality class.
\end{abstract}


\maketitle

\section{Introduction}

Growing surfaces in two dimensions which break time-reversal but not rotational or translational symmetry are known to fall in the Kardar-Parisi-Zhang (KPZ) universality class \cite{1kpz,1krapivskybook}. This means that the exponents governing the long-time and long-distance behaviour of single point fluctuations and two-point correlations are same for the models within this class. However, specific models belonging to this class have been shown to have a deeper connection \cite{baik,1kriecherbauerkrug,1satyakpzrev}. Examples of problems shown to have this connection are the TASEP on an infinite lattice, Anisotropic Directed Percolation in 3D \cite{1rajeshdhar,1seppalainen}, the Bernoulli Matching problem \cite{1bmfirst,1bmsatya1,1bmsatya2}, Random Growth models \cite{ferrarispohn}, and the largest eigenvalue problem in Random Matrix theory \cite{deansatya,satyarm}. In these problems, one is typically concerned with the distribution of fluctuations of a random variable $h(t)$ (which could be the height profile in a random growth model), and remarkably, in all these cases this distribution has the form
\beq 
P(h,t) = A t^{2/3} U_{\beta} (B h/t^{2/3}),
\label{eq:tw}
\eeq
where $U_{\beta}$ is the Tracy-Widom distribution with $\beta=1$, $2$ or $4$  \cite{1tracywidom,1priezzhevbm,1johansson02,satyaderiv} ($A$ and $B$ are model-dependent constants). The growing evidence for Tracy-Widom distributions occuring in seemingly unrelated non-equilibrium phenomena points to an new underlying universality that is yet to be fully understood. There has been an explosion of interest in Tracy-Widom distributions amongst theoretical and experimental physicists\cite{halpin12,twliquidcrys,takeuchisr}, mathematicians\cite{twmathrev}, and biologists\cite{twgenomics}.

\begin{figure}[b]
	\centering
	\includegraphics[width=0.9\columnwidth]{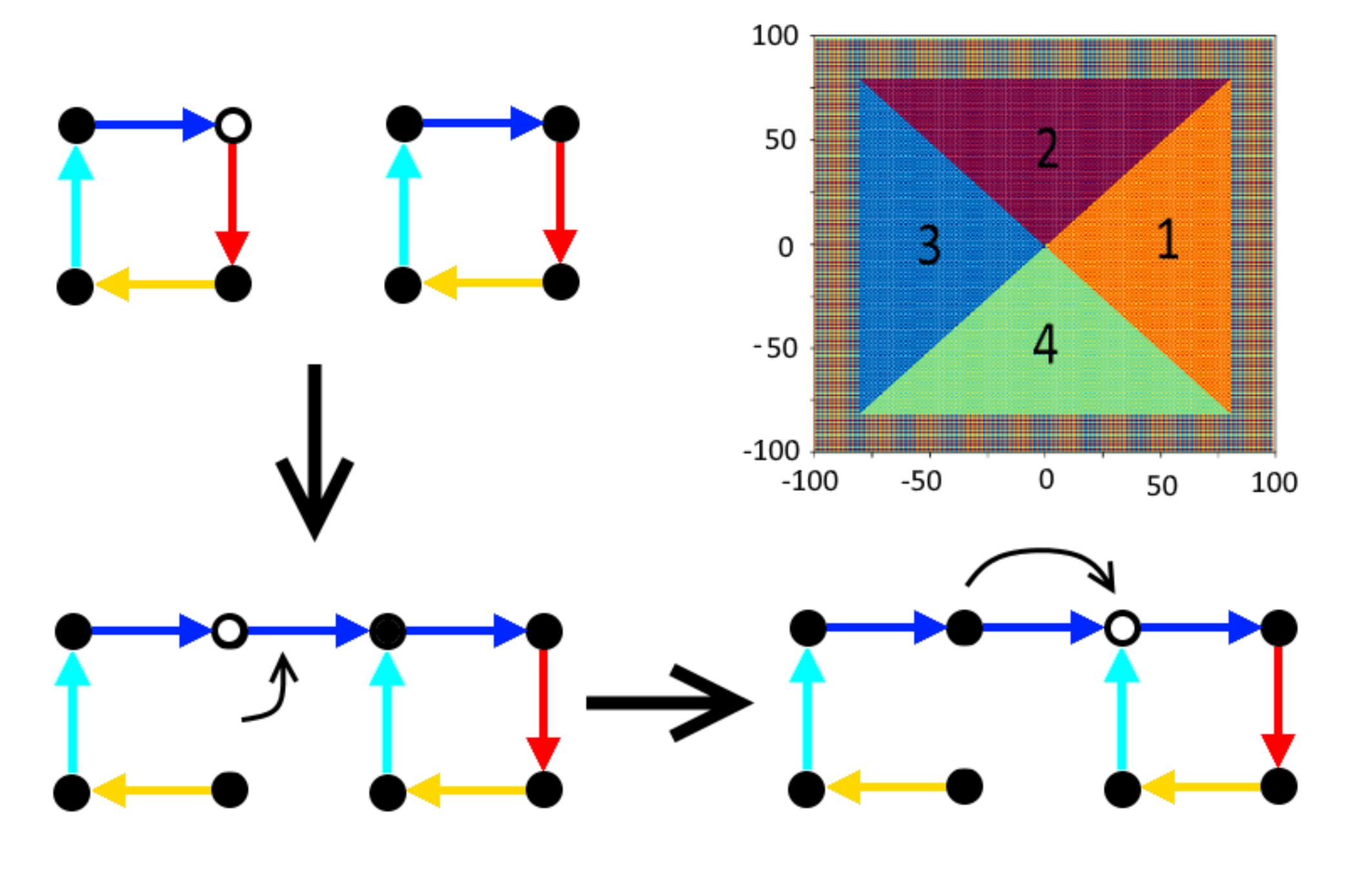}
	\caption{The rotor-router walker (whose position is indicated by the open circle) turns an arrow counterclockwise and steps to the next site along the new arrow direction. Inset: Pattern formed on the Type I background with $p=0$, after the first walker is stopped after visiting the origin 180 times. The patches are numbered for convenience in referring to them in the text. Colour code: dark blue - $\rightarrow \equiv 0$, light blue - $\uparrow \equiv 1$, yellow - $\leftarrow \equiv 2$, red - $\downarrow \equiv 3$.}
	\label{rrmove}
\end{figure}

Many systems in nature are believed to spontaneously reach a critical state, a phenomenon known as self-organized criticality (SOC) \cite{bakbook}. Examples of simple models which show self-organized criticality are the sandpile model \cite{btw,dharbtw} and the rotor-router model \cite{1pddk}. In recent years, some of these models have been snown to exhibit intricate growing patterns which show proportionate growth \cite{1dsc09,1sd12}, where all parts of the pattern, whether big or small, grow at the same rate, as happens in biological systems such a mammalian bodies, but without a pre-planned program for growth or co-ordination. However, in the case of the sandpile model, noise in the initial condition eventually destroys the intricate structure of the pattern \cite{1sdjstat}. In this paper we study the patterns formed in the rotor-router model, which was introduced as the Eulerian Walkers model in the physics literature \cite{1pddk,1spp97,1pps98}. The intricately structured patterns observed in this model, when a pattern is grown on periodic backgrounds, have been studied previously \cite{1dandekardhar}. Meanwhile, numerical studies on completely random backgrounds have shown that the boundary of the cluster of visited sites displays KPZ fluctuations \cite{1kapridhar,priezzrr1,priezzrr2}, although there is little discernible patterning in the bulk. In this Letter, we study the effect of a controllable amount of initial (quenched) noise on the pattern, and show that the pattern is not destroyed even for a finite amount of noise. In particular, we provide an exact mapping from the patterns formed by a rotor-router walker on a lattice with quenched disorder to height models in the Tracy-Widom class, thereby showing that the Tracy-Widom universality extends to deterministic cellular automata with quenched randomness. 

The properties of the rotor-router are intimately related to the properties of the random walk on the corresponding lattice \cite{1propprev,1mathreview,1levinethesis,1levineperes2}. We study the rotor-router model in 2D, the critical dimension at which random walks change between recurrence and transience \cite{doylesnell}. We show for the first time that the 2D rotor-router exhibits a transition between transience and recurrence. The classification of irreducible Markov chains as recurrent or transient is an important and fundamental problem\cite{markov}. Recently, there has been a surge of interest in the recurrent and transient properties of de-randomized, deterministic versions of Markov chains, of which the rotor-router model is one example \cite{rr1,rr2,rr3,rr4,rr5,1holpropp,rr6,rr7,rr8}. We show that a recurrence-transience transition can be produced by tuning the amount of quenched disorder, and that this transition falls in the well-known 3D anisotropic directed percolation universality class. We provide both analytic arguments and extensive numerical evidence to support our claims.\\

\textsl{The rotor-router model:} We consider the rotor-router model on a two dimensional square lattice. Each site on the lattice is equipped with a single arrow that points to one of the neighbouring sites (see Fig. \ref{rrmove}). The path of a walker on this lattice is directed by the arrows, and in turn changes their orientation: the rule is that a walker, when it visits a site, first rotates the arrow attached to the site by $90^{\circ}$ counter-clockwise, and then walks to the next site along the new direction of the arrow. Starting from a configuration where every arrow points to the right, a walker inserted at the origin would walk straight along the $y$-axis to infinity, changing all arrows along the $y$-axis to point up. Such configurations, on which the walker visits each site only finitely many times, are called transient backgrounds. On arrow configurations where this does not happen, the walker visits every site on the lattice infinitely many times before it reaches infinity \cite{1holpropp}.

A simple example of a recurrent configuration is the periodic configuration constructed by tiling the lattice with the $2 \times 2$ unit cell $\left( \besm \rightarrow & \downarrow \\ \uparrow & \leftarrow \eesm \right)$, also written as $\left( \besm 0 & 3 \\ 1 & 2 \eesm \right)$ with the notation $0,1,2,3$ for arrows pointing right, up, left and down respectively. Throughout the paper, we will study the pattern created by a single walker introduced at the origin (the centre of the lattice). The growing pattern that such a walker makes on the recurrent background given above consists of four patches, as shown in the inset of Fig. \ref{rrmove}. The path of the walker is characterized by the Visit (or Odometer) function $V(i,j,M)$ \cite{1propprev,1mathreview}, which, for a recurrent background, counts the number of times the site $(i,j)$ has been visited by the walker before it reaches the origin $M$ times. For a transient background, since the origin is only visited a finite number of times, the analogous function is $V_{tr}(i,j)$, which counts the total number of times the site $(i,j)$ is visited before the walker reaches infinity. On the recurrent background created by tiling the lattice with $\left( \besm 0 & 3 \\ 1 & 2 \eesm \right)$, it is easy to show that the visit function is given by 
\beqa
V^0(i,j) = 
\begin{cases}
	M - 2\lvert i\rvert & \text{in patches 1 and 3},\\
	M - 2\lvert j\rvert & \text{in patches 2 and 4},
\end{cases}
\label{eq:V0eq}
\eeqa

\noindent while $V^0(i,j)=0$ outside the pattern, that is outside the square of diameter $M$ centred on the origin. The number of visits to the origin increases linearly with the diameter of the pattern. We denote arrow configurations by the notation $\rho(i,j)$, where $\rho = 0,1,2,3$. Given $V(i,j,M)$ and the initial arrow configuration $\rho_i(i,j)$, the final arrow configuration is given by $\rho_f = \lfloor(\rho_i+V)/4\rfloor$. For the rotor-router model, $V(i,j,M)$ function obeys the Laplace-like equation
\beq
\mathcal{L}~ [V,\rho_i] (x,y) \equiv \nabla^2 V(x,y) + \sum 
f(x',y',x,y) = 0 
\label{eq:LV}
\eeq
\noindent where the summation is over the neighbours of site $(x,y)$. The function $f_{(\rho_i,\rho_f)}(x',y',x,y)$ can take the value $0$ or $1$, and it counts the number of times the arrow at site $(x',y')$ points towards site $(x,y)$ when one rotates it from $\rho_i(x',y')$ to $\rho_j(x',y')$.\\

\begin{figure}[t]
	\centering
	\includegraphics[width=0.95\columnwidth,height=0.78\columnwidth]{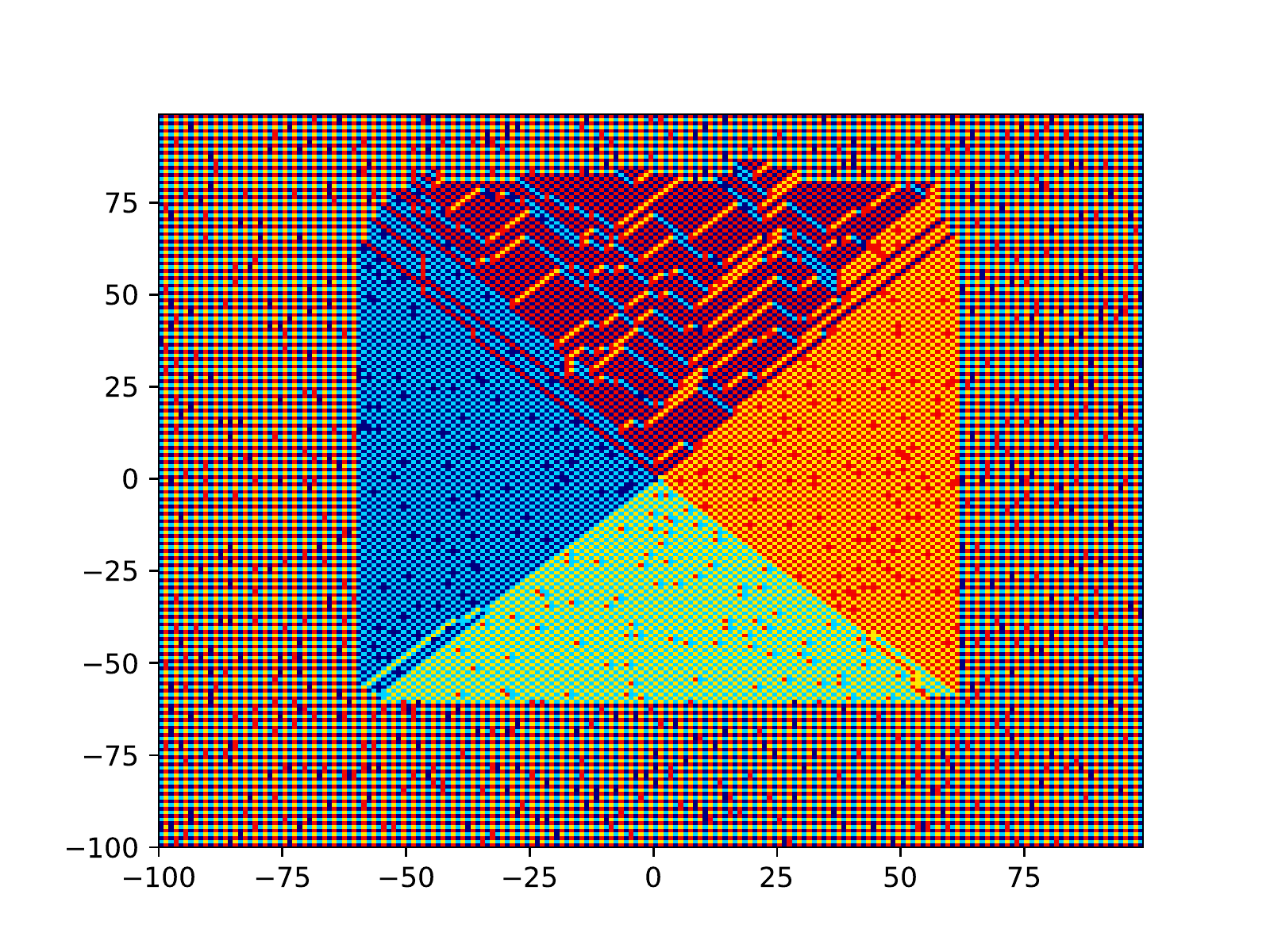}
	\caption{Pattern formed on with a type I perturbation, with $p=0.05$. Colour code is as in Fig. \ref{rrmove}.}
	\label{4big1}
\end{figure}

\textsl{Arrow configurations with quenched randomness:} We construct random arrow configurations by replacing some of the unit cells in the periodic configuration above by a certain fraction of `defect cells'. The fraction of defects controls the amount of randomness in the pattern.  We first study the perturbation of the periodic initial condition defined above, such that a fraction $p$ of all arrows on odd sites are turned counterclockwise once (this we call the `type I' background). The probabilities of different unit cells are then
\beqa
\textmd{Prob} \left[\left( \besm 0 & 3 \\ 1 & 2 \eesm \right)\right] = (1-p)^2 ~&,&~ \textmd{Prob}\left[ \left( \besm 0 & 3 \\ 1 & 3 \eesm \right) \right] = p(1-p), \nonumber \\ 
\textmd{Prob}\left[ \left( \besm 0 & 3 \\ 0 & 2 \eesm \right)\right] = p(1-p) ~&,&~ \textmd{Prob}\left[ \left( \besm 0 & 3 \\ 0 & 3 \eesm \right) \right] = p^2,
\label{eq:type1}
\eeqa
with $p$ varying between 0 to 1. $p=0$ is the unperturbed background, while at $p=1$ the i.c. is a perfect tiling of the lattice by the unit cell $\left( \besm 0 & 3 \\ 0 & 3 \eesm \right)$, which is a transient background, with a particle inserted at the origin moving in a straight line along the positive y-axis and leaving the lattice. The randomness in the initial condition has a local effect in patches 1, 3 and 4, but creates terrace-likes structures in patch 2, as seen in Fig. \ref{4big1}.  We shall focus on the effect in patch 2.

\begin{figure}[t]
	\centering
	\includegraphics[width=0.9\columnwidth]{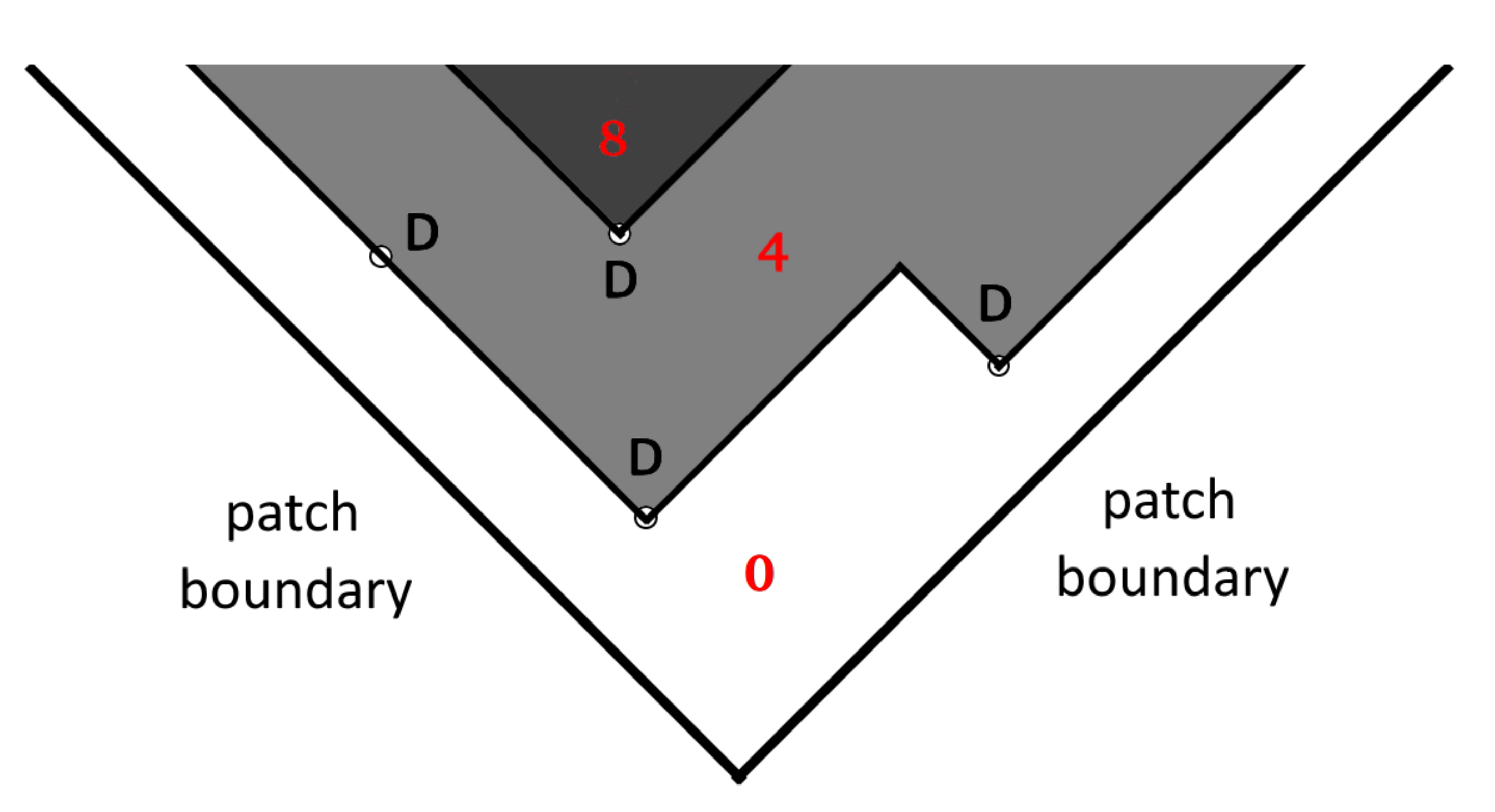}
	\caption{The rules for constructing the terraces in patch 2, based on the positions of the perturbed unit cells (defects, indicated by `D'). The numbers show the height field $h(x,y)$, which is constant within a given terrace, but changes by $4$ when moving from the outside of a terrace to the inside.}
	\label{hrules}
\end{figure}

We now show that the structure in patch 2 can be mapped to an exactly solvable height model. Eq. (\ref{eq:LV}), along with the fact that the final arrow configuration should be loopless \cite{1pps98}, and the condition $V(0,0,M)=M$ can be used to generate a solution $V(i,j,M)$ for the Visit function, starting from an initial guess $V^0(i,j)$. (See SI for details \cite{si} for the algorithm. A similar algorithm was earlier used to study rotor-router aggregation \cite{1levinefast}.) We start from the initial guess $V^0(i,j)$, Eq. (\ref{eq:V0eq}), which is the correct Visit function for $p=0$, and determine the effect of adding randomness to the pattern. We prove that (for details and proof, see SI \cite{si}) the true visit function $V(i,j,M)$ can be written as  $V_0(i,j,M) + h(i,j,M)$. We call $h(i,j,M)$ the `height field', and it can be determined from the positions of the perturbed unit cells, by a few simple rules: 

(1) A single perturbed unit cell creates a terrace inside which the $h$ is increased by $4$ compared to the $h$ outside the cone. Thus two nested terraces will successively increase $h$ by 4 when crossing from the outside of both to the inside of both (See fig. \ref{hrules}).

(2) Two intersecting `V'-shaped terraces merge to create a single `W'-shaped terrace. 

(3) A perturbed unit cell lying on the boundary of a terrace does not have any effect.\\

The terraces constructed using these rules can be used to determine $h(i,j)$, and hence $V(i,j)$, for all $p$. The rules given above are in fact the same as those used to create terraces of wetted sites in 3D anisotropic directed percolation (ADP) \cite{1rajeshdhar,1seppalainen}, which can also be exactly mapped to the Bernoulli matching problem and the longest increasing subsequence problem \cite{1satyakpzrev}. The exact solution for the height field is known in this case, and thus we can write down that $h(i,j) = 4 h_{BM}(\frac{i-j}{2},\frac{i+j}{2})$ (as the patch is aligned with the $j$-axis as its diagonal, and the terrace boundaries have a width of 2 lattice units), where $h_{BM}$ is the height field in the Bernoulli Matching problem. Hence \cite{1priezzhevbm},

\beqa
\frac{h(i,j)}{4} = 
\begin{cases}
	j-i & \text{if } \frac{i}{j} < -\frac{1-p}{1+p},\\
	i+j & \text{if } \frac{i}{j} > \frac{1-p}{1+p}, \\
	\frac{\sqrt{p (j^2-i^2)} -  p j}{1-p} + A(p,i,j) \xi_{GUE} & \text{otherwise}
\end{cases}
\label{eq:4hBM}
\eeqa

\begin{figure}[t]
	\centering
	\includegraphics[width=0.9\columnwidth,height=0.65\columnwidth]{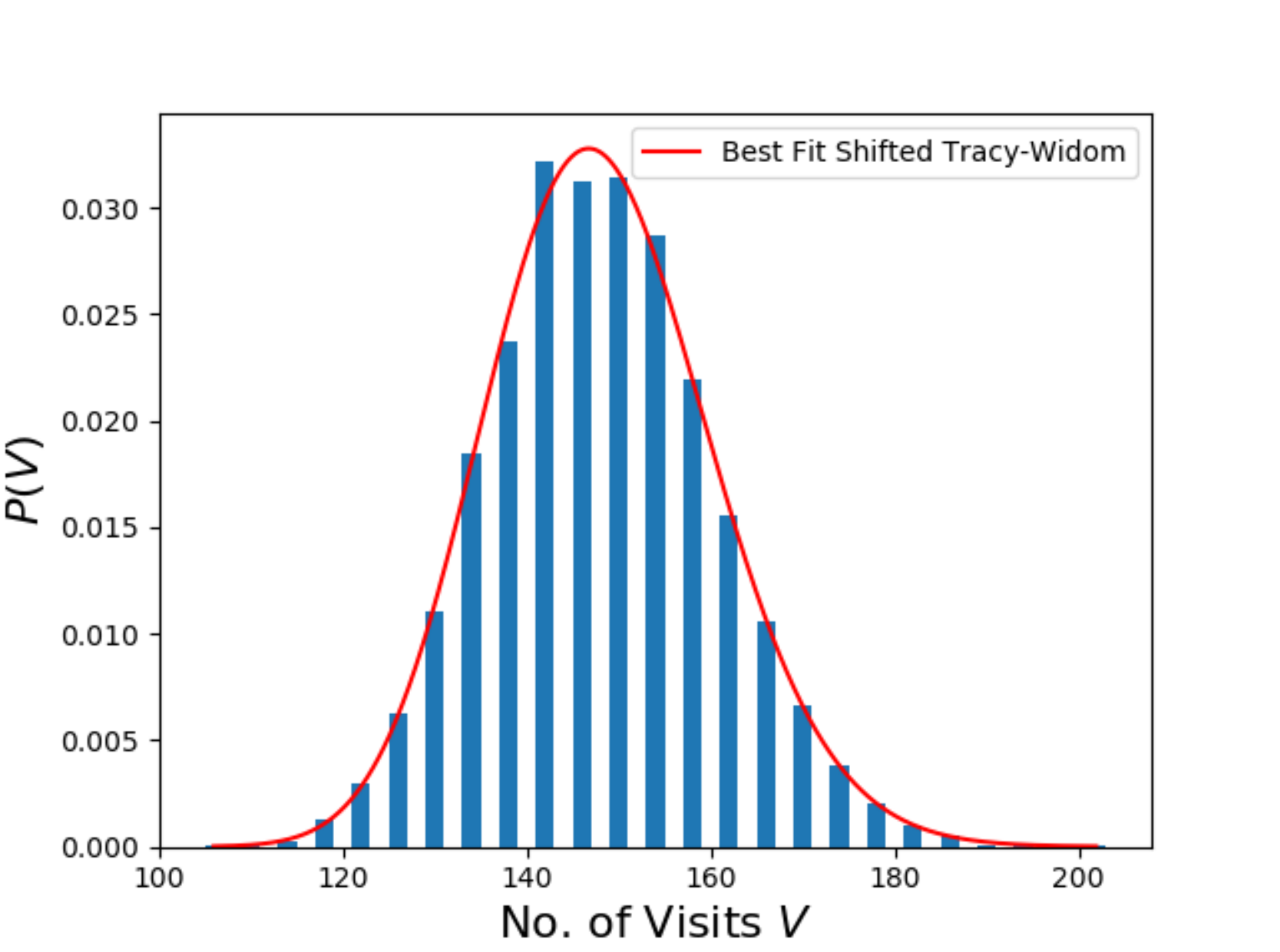}
	\caption{The histogram for fluctuations about the average height for a type I pattern with $p=0.6$, at a point at $(i,j) = (0,2480)$ along the y-axis, with $V(0,0)=1200$. Also shown is the best fit shifted Tracy-Widom type-2 distribution, $TW_2(\xi,m,s)$. The best fit values are $m=172.16$ and $s=13.57$. From eqn. \ref{eq:4hBM}, we get the theoretical predictions $m_{th} = 170.00$ and $s_{th} = 13.5346$.}
	\label{tw1}
\end{figure}

\noindent where $\xi_{GUE}$ is a random variable distributed according to the Tracy-Widom distribution corresponding to the Gaussian Unitary Ensemble, also known as the Tracy-Widom Type 2 distribution, and $A(p,i,j) = \frac{(p(j^2-i^2))^{1/6}}{2^{1/3} (1-p)} \left[ (1+p) - 2 \sqrt{\frac{p}{(j^2-i^2)}} j \right]^{2/3}$.\\

To compare with simulations, we define the shifted Tracy-Widom type 2 distribution $TW_2(\xi_s,m,s)$ as the distribution followed by the variable $\xi_s = m + s \xi_{GUE}$. Thus,
\beq
TW_2(\xi_s,m,s) = U_{\beta=2} \left(\frac{\xi_s-m}{s}\right)
\eeq
where $U_{\beta=2}$ is the Tracy-Widom distribution with $\beta=2$. Fig. \ref{tw1} shows that the measured distribution of height fluctuations in a type I pattern (within patch 2) agrees extremely well with the prediction in Eqn. (\ref{eq:4hBM}). Eqn. (\ref{eq:4hBM}) also shows that the background in Eq. (\ref{eq:type1}) is transient only at $p=1$. Thus, a transition from recurrence to transience happens at $p=1$, and it falls in the universality class of the wetting transition seen in 3D anisotropic directed percolation \cite{1rajeshdhar}.\\

\begin{figure}[b]
	\centering
	\includegraphics[width=0.9\columnwidth,height=0.82\columnwidth]{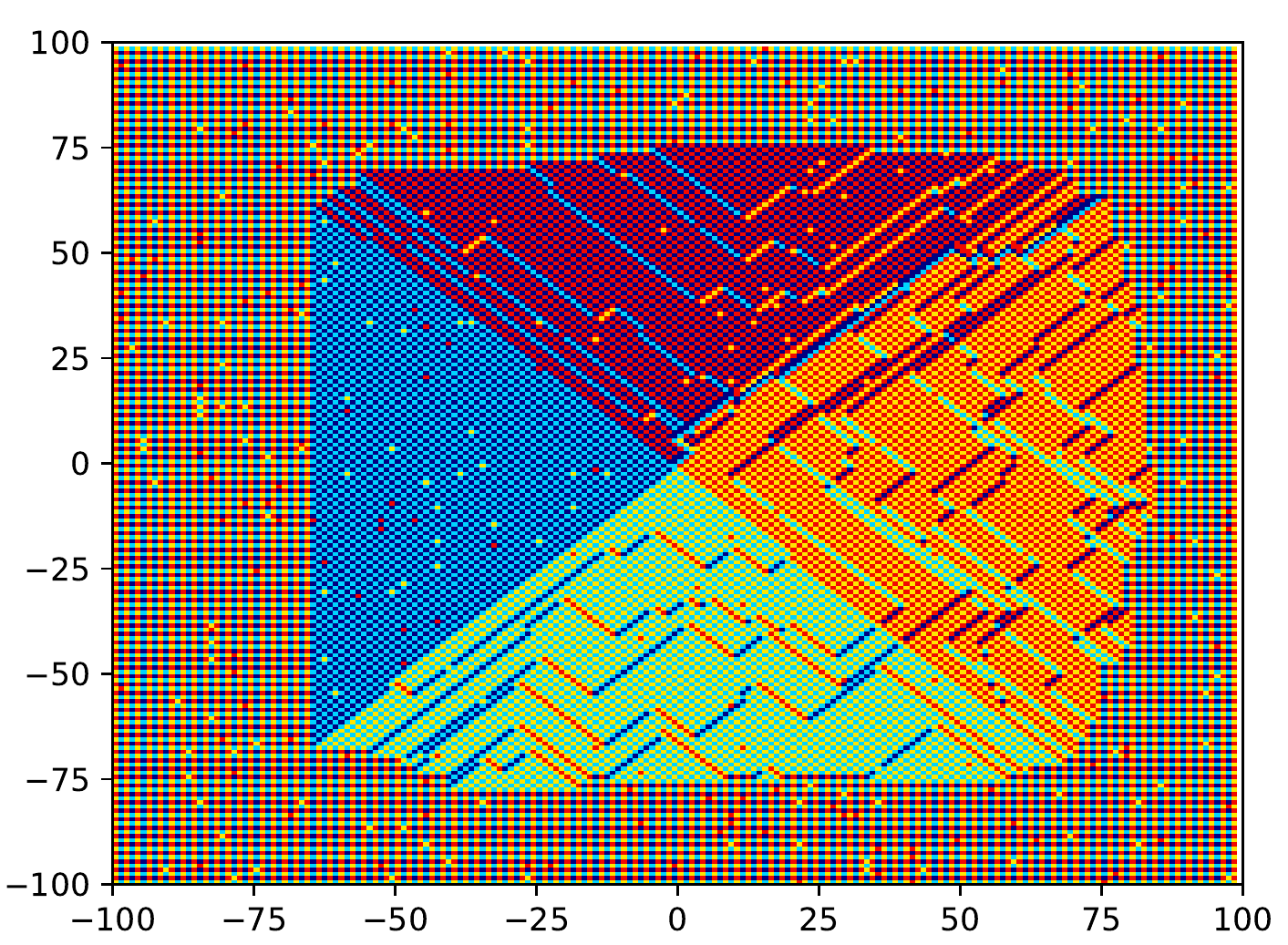}
	\caption{Pattern formed on a type II background, with $p=0.02$. The pattern shows terraces in three patches. Colour code as in Fig. \ref{rrmove}.} 
	\label{4big2}
\end{figure}

\begin{figure}[t]
	\centering
	\includegraphics[width=0.9\columnwidth]{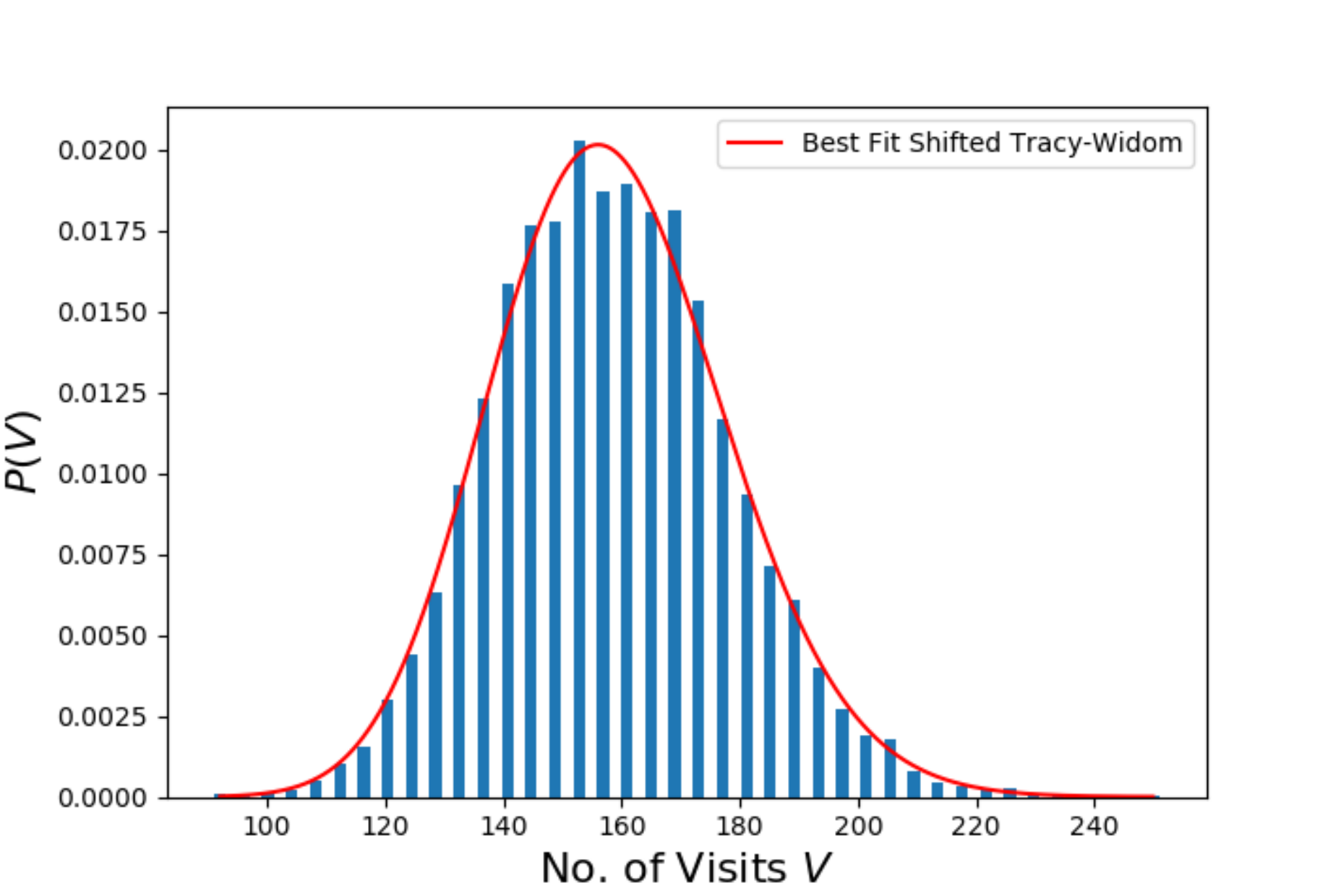}
	\caption{The histogram for fluctuations about the average height for a type II pattern with $p=0.2$. Also shown is the best fit shifted Tracy-Widom type-2 distribution, with the best fit values $m=197.32$ and $s=22.08$. Since this is only the central part of the distribution, the fits for shifted Tracy-Widom distributions with $\beta=1$ and $4$ are equally good, but with different parameter values (not shown).}
	\label{tw2}
\end{figure}

\begin{figure}[t]
	\centering
	\includegraphics[width=0.9\columnwidth]{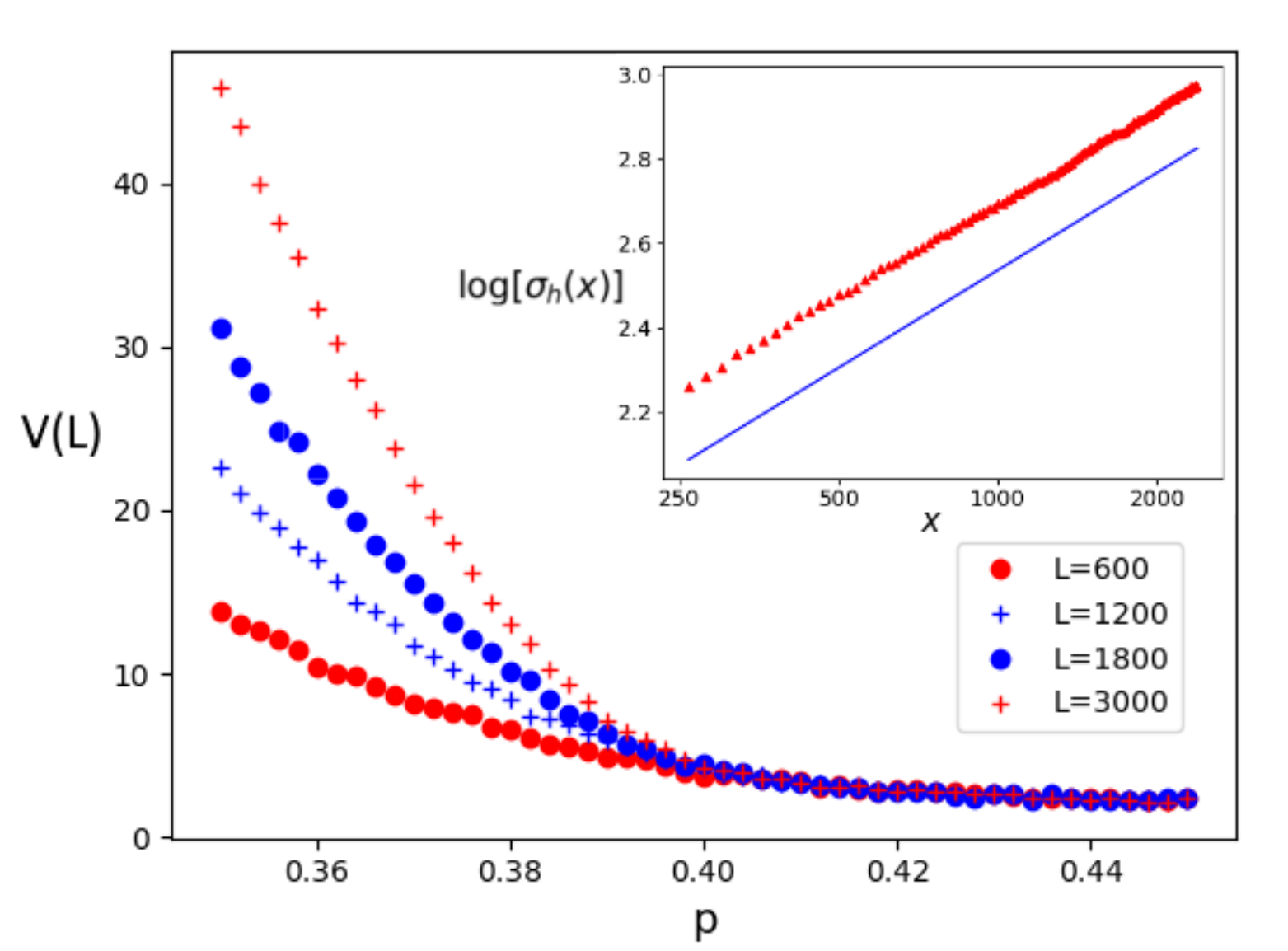}
	\caption{$V(L)$ vs. $p$ for the type II pattern. $V(L)$ is the number of visits to the origin before the walker reaches a site with $x=L$, which increases with $L$ in the recurrent regime (below the transition), but is constant in the transient regime. Inset: Log-log plot of the mean squared fluctuations in height along the $x$-axis, against $x$. The blue line shows the best fit slope $0.33$.}
	\label{type2trans}
\end{figure}

\textit{A background with a transition at $p<1$:} We next study the recurrence-transience transition in detail for a case that cannot be exactly mapped to a solvable model. Consider the pattern formed by the walker on the random arrow configuration constructed out of 2$ \times $2 unit cells with the following probabilities (`type II'):
\beqa
\textmd{Prob}\left(\left( \besm 0 & 3 \\ 1 & 2 \eesm \right)\right) = (1-p)^2 ~&,&~ \textmd{Prob}\left( \left( \besm 2 & 3 \\ 1 & 2 \eesm \right) \right) = p(1-p), \nonumber \\ 
\textmd{Prob}\left(\left( \besm 0 & 3 \\ 3 & 2 \eesm \right)\right) = p(1-p) ~&,&~ \textmd{Prob}\left( \left( \besm 2 & 3 \\ 3 & 3 \eesm \right) \right) = p^2.
\label{eq:type2}
\eeqa
For $p=0$, this is the unperturbed background whose visit function is given by Eq. (\ref{eq:V0eq}), while for $p=1$ it is transient. Fig. \ref{4big2} shows that the defects create terraces in three patches of the unperturbed background. The recurrence-transience transition happens at $p_c \approx 0.4$ (see Fig. \ref{type2trans}). The rules for constructing the terraces can be determined, and are not the same as for the previous case, and cannot be exactly mapped to a solvable model (see SI \cite{si}). But the terraces and the corresponding height field $h(i,j) = V(i,j) - V^0(i,j)$ can still be constructed given the positions of the defect cells. The height increases by $4$ when crossing from the outside of a terrace to the inside. It can be seen that the height-fluctuations follow a Teacy-Widom distribution in this case as well \ref{tw2}, although the data in the tails is not good enough to distinguish between the cases $\beta=1$, $2$ and $4$.

We next give an interpretation of the transition in terms of the height field. Since the boundary of the pattern is given by $V^0(i,j,M)+h(i,j)=0$, the transition point is simply the point at which the (positive) slope of the height field equals the (negative) slope of $V^0$, which is $2$ (see Eq. \ref{eq:V0eq}). For type II patterns, the height field increases fastest along the positive x-axis, and thus, the recurrence-transience transition at $p=p_c$ happens along a preferred direction, the positive x-axis. Above $p=p_c$ the height field calculated from the constructed terraces does not describe the path of the walker, as $V = V^0 + h$ has a positive slope and thus cannot be the solution of Eq. (\ref{eq:LV}). However, nothing singular happens to the height field itself at $p=p_c$. Even though an exact mapping to a solvable model cannot be constructed for type-II patterns, simulations show that the height field $h(i,j)$, and hence also the boundary of the pattern, shows fluctuations which increase as $\langle h \rangle^{1/3}$, or as $R^{1/3}$ where $R$ is the distance from the origin (see Fig. \ref{type2trans}, inset).

Properties of the transition can be determined by looking at the height field. Below $p_c$, the probability that the pattern extends a distance $L$ is, for large $L$, a decaying function of the quantity $L/\xi(p)$, where $\xi(p)$ is a length scale that diverges as $(p-p_c)^{-\nu}$, defining the correlation length exponent $\nu$. We now calculate the exponent $\nu$. As mentioned earlier, the height field does not exhibit any discontinuity at $p_c$, only that its slope along the x-axis crosses the value $2 \equiv m_0$. Let the slope of the height field along the x-axis be $m = m_0 - \delta m$ for $p = p_c - \delta p$, where $\delta m \sim \delta p$, assuming nothing singular happens to the height field at $p=p_c$. At a distance $L$ from the origin, thus, $\langle V(L) \rangle = - L (\delta m)$. For this site to be part of the pattern, $V(L)>0$. The variable $V(L)$ fluctuates about its average with variance $L^{1/3}$. The probability that one observes a fluctuation of order $\langle V(L) \rangle$ is thus a decaying function of $\langle V(L) \rangle/ L^{1/3}$, and hence of the variable $L (\delta m)^{3/2} \sim L (\delta p)^{3/2}$, giving $\nu = 3/2$. 

The same correlation length exponents are found for the wetting transition in the 3D Anisotropic Directed Percolation model \cite{1rajeshdhar}, along with KPZ fluctuations of the height field. This strongly suggests that the recurrence-transience transition in rotor-routers belongs to the 3D ADP class.\\

\textsl{Conclusions and Outlook:} We have shown that, for some types of quenched randomness, there is an exact mapping of the pattern formed by the rotor-router walker to the Anisotropic Directed Percolation model. This allowed us to calculate the distribution of the number of visits to a site, which follows a Tracy-Widom Type-II distribution. We also found that for another case where an exact mapping cannot be performed, a height representation exists below $p_c$, and the pattern also shows KPZ fluctuations. The properties of the recurrence-transience transition can be described by the wetting transition in 3D ADP. The properties of the transient walk above $p_c$ are yet to be elucidated.

Although the KPZ property can only be proved in certain cases, we have shown that it holds for a wider range of recurrent backgrounds. For the cases when the recurrence-to-transience transition happens along a preferred direction, and the height field in that direction shows KPZ fluctuations, the analysis performed for type II patterns can be generalized, and thus the transition should be in the ADP universality class. Also of interest is the effect of quenched randomness in the initial configuration on the intricately structured patterns created by a large number of walkers starting at the origin on periodic transient backgrounds\cite{1dandekardhar}. Similar patterns are also seen when the initial conditions of linearly growing sandpiles \cite{1sd12} are perturbed \cite{sdpriv}. It would be interesting to extend the KPZ analysis to this case.


\begin{acknowledgments}
I would like to thank Prof. Deepak Dhar for his constant guidance and many helpful discussions, and Kabir Ramola for helpful comments on the manuscript.
\end{acknowledgments}

\end{document}


\maketitle

In this supplementary information, we provide proofs and results for certain assertions made in the main paper.\\

First, we study the background of type I, and prove the rules for the terraces and thus the height field for this pattern, thus proving the mapping to the Anisotropic Directed Percolation model and related models.\\

Then, we give the rules for constructing the terraces of backgrounds of type II. We show simulation results supporting the assertion that the height field in this case shows KPZ fluctuations.\\

\section{Type I backgrounds}

To recap, the type I background is given by the equation

\beqa
\textmd{Prob} \left[\left( \besm 0 & 3 \\ 1 & 2 \eesm \right)\right] = (1-p)^2 ~&,&~ \textmd{Prob}\left[ \left( \besm 0 & 3 \\ 1 & 3 \eesm \right) \right] = p(1-p), \nonumber \\ 
\textmd{Prob}\left[ \left( \besm 0 & 3 \\ 0 & 2 \eesm \right)\right] = p(1-p) ~&,&~ \textmd{Prob}\left[ \left( \besm 0 & 3 \\ 0 & 3 \eesm \right) \right] = p^2,
\label{eq:type1}
\eeqa

The unit cells $\left( \besm 0 & 3 \\ 1 & 2 \eesm \right)$, $\left( \besm 0 & 3 \\ 1 & 3 \eesm \right)$, $\left( \besm 0 & 3 \\ 0 & 2 \eesm \right)$ and $\left( \besm 0 & 3 \\ 0 & 3 \eesm \right)$ form the type I background, being present with probabilities $(1-p)^2$, $p(1-p)$, $p(1-p)$ and $p^2$ respectively.

After a mapping to the Bernoulli Matching problem, discussed below, we have that the height field for this background is given by $h(i,j) = 4 h_{BM}(\frac{i-j}{2},\frac{i+j}{2})$, and hence
\beqa
\frac{h(i,j)}{4} = 
\begin{cases}
	j-i & \text{if } \frac{i}{j} < -\frac{1-p}{1+p},\\
	i+j & \text{if } \frac{i}{j} > \frac{1-p}{1+p}, \\
	\frac{\sqrt{p (j^2-i^2)} -  p j}{1-p} + A(p,i,j) \xi_{GUE} & \text{otherwise}
\end{cases}
\label{eq:4hBM}
\eeqa

\begin{figure}[t]
	\centering
	\subfigure{
		\includegraphics[width=0.45\columnwidth]{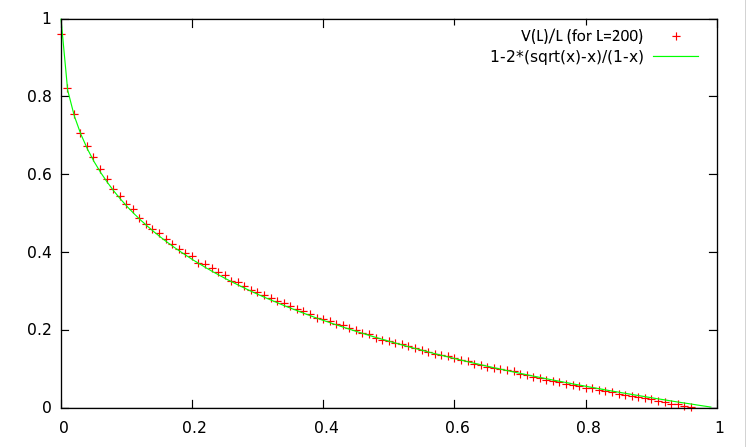}
	}
	\subfigure{
		\includegraphics[width=0.45\columnwidth]{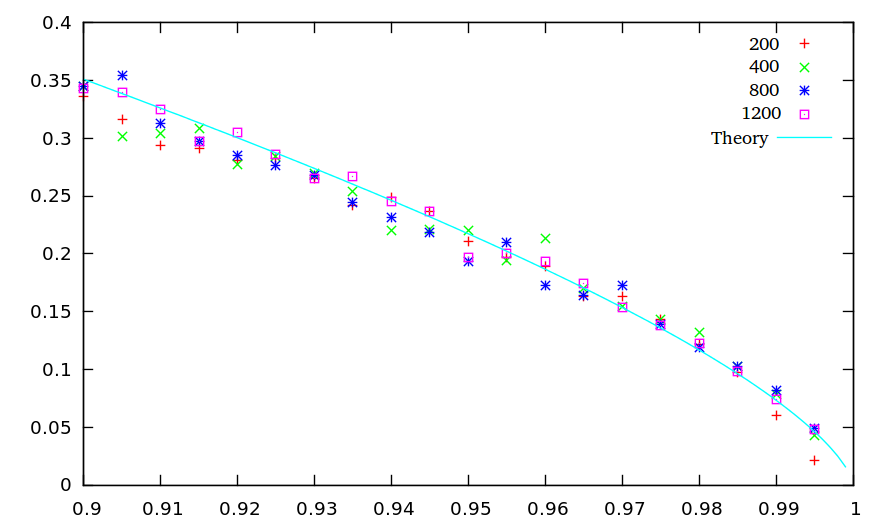}
	}
	\caption{(a) Number of visits to origin$/L$ vs. $p$: Comparison of theory and simulation results for type (i) backgrounds. (b) $\sigma_{V(0)/L} \times L^{-1/3}$ vs. $p$: Comparison of theory and simulation for scaled variance of order parameter.}
	\label{type1trans}
\end{figure}

We compare this with simulations by using the equation $h(0,L)+V^0(0,L,M)=0$ to find the mean and variance of $M$, the number of visits to the origin, when the boundary of the pattern is at a distance $y=L$. Fig. \ref{type1trans} shows that the simulation data are in good agreement with the prediction from Eq. (\ref{eq:4hBM}).\\

Now we prove the mapping to the Bernoulli matching problem, using an algorithm for generating $V(i,j,M)$ starting from an initial guess $V^0(i,j,M)$.

\subsection{Constructing the terraces from studying point pertubations}

The Visit function for the rotor-router obeys the equation 
\beq
\mathcal{L}~ [V,\rho_i] (x,y) \equiv \nabla^2 V(x,y) + \sum 
f(x',y',x,y) = 0. 
\label{eq:LV}
\eeq
This equation, along with the fact that the final arrow configuration should be loopless, allows one to use an algorithmic procedure to generate $V(i,j,M)$. A similar algorithm was earlier used by Friedrich and Levine to study the patterns formed in Rotor-router aggregation \cite{4levinefast}. The algorithm consists of the following steps:
(a) Start with a guess $V^0(i,j,M)$. Evaluate the final arrow configuration and $\mathcal{L} V$.\\
(b) The sites where $\mathcal{L} [V]>0$ are called hills and the sites where $\mathcal{L} [V]<0$ are called holes. Place walkers at each hill and allow them to walk until they reach a hole, changing the odometer function appropriately.\\
(c) Scan the lattice for loops formed by arrows. Follow a procedure called reverse-cycle-popping, which gets rid of these loops.\\
(d) Steps (b) and (c) give an updated guess for the Visit function as follows: extra walkers add to $V$, while reverse-cycle-popping reduces $V$ on the sites along the loop. With this updated guess, go back to (a).\\

If a loop, or `cycle', exists in an arrow configuration,`reverse cycle popping' is the operation where all the arrows which form the cycle are rotated $90^{\circ}$ clockwise (that is, reverse of the anti-clockwise direction in which they are moved by the walkers). This destroys the original loop, but might create new loops. It is useful to extend the definition of reverse cycle popping to a set of arrows which do {\emph not} form a loop as simply doing nothing to the arrow configuration. Friedrich and Levine showed that the procedure of reverse cycle-popping a set of sites commutes with the operation of starting a walker at a site and letting it walk until it reaches the sink. That is, reverse-popping the loop before starting the walk will result in the same final configuration once the walker has reached the sink as performing the reverse-popping after the walker has reached the sink - if the loop still exists, that is. (If the walker visits any site on the loop, the final configuration will not contain that loop, and reverse-cycle popping will do nothing.)\\

\begin{figure}[h]
	\centering
	\subfigure[]{
		\includegraphics[width=0.46\linewidth,height=0.44\linewidth]{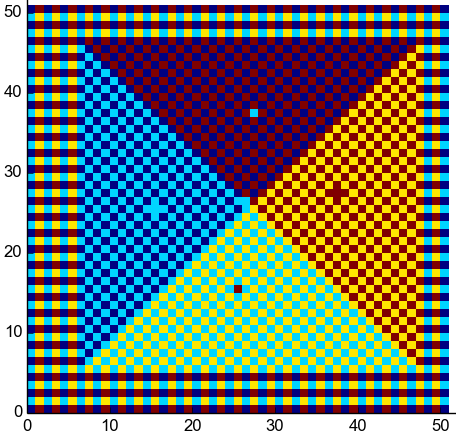}
	}
	\subfigure[]{
		\includegraphics[width=0.46\linewidth,height=0.44\linewidth]{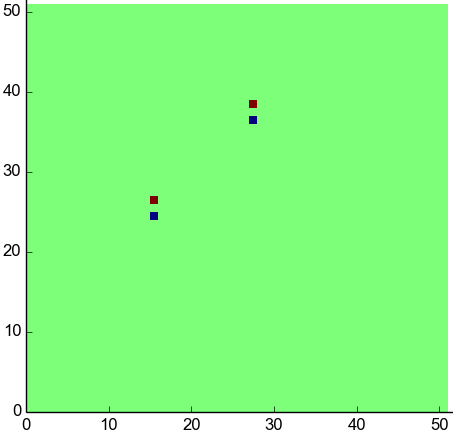}
	}
	\caption{(a) The initial approximation to the final arrow configuration. The differently colored cells in the four patches are where the arrow is flipped from the periodic background. Colour code: dark blue - $\rightarrow$, light blue - $\uparrow$, yellow - $\leftarrow$, red - $\downarrow$ (b) $\mathcal{L}~u_0(i,j)$ for initial approximate visit function $u_0(i,j)$. Color code: red = 1, blue = -1, green  = 0.}
	\label{4type1-fl}
\end{figure}

To use the above procedure, one first considers point perturbations. We call a perturbation in which only a single unit cell of a periodic background is changed a `point perturbation'. Consider a point perturbation of the background $\left( \besm 0 & 3 \\ 1 & 2 \eesm \right)$ such that only a single arrow is rotated $90^{\circ}$ anticlockwise, changing a single unit cell in the pattern from  $\left( \besm 0 & 3 \\ 1 & 2 \eesm \right)$ to  $\left( \besm 0 & 3 \\ 1 & 3 \eesm \right)$. We want to determine the final configuration we get after a single walker started at the origin walks until it has visited the origin $N$ times, where $N$ is such that the region visited by the walker contains the perturbed site. For the case shown in fig \ref{4type1-fl}, a four unit cells of the initial background are changed in the above fashion, and $N$ is $40$.\\

\begin{figure}[h]
	\centering
	\subfigure[]{
		\includegraphics[width=0.46\linewidth,height=0.44\linewidth]{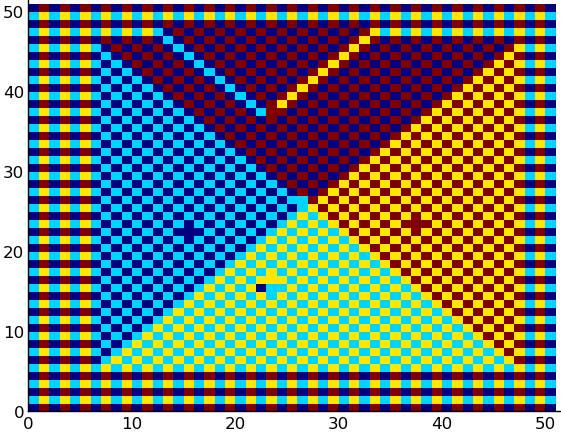}
	}
	\subfigure[]{
		\includegraphics[width=0.46\linewidth,height=0.44\linewidth]{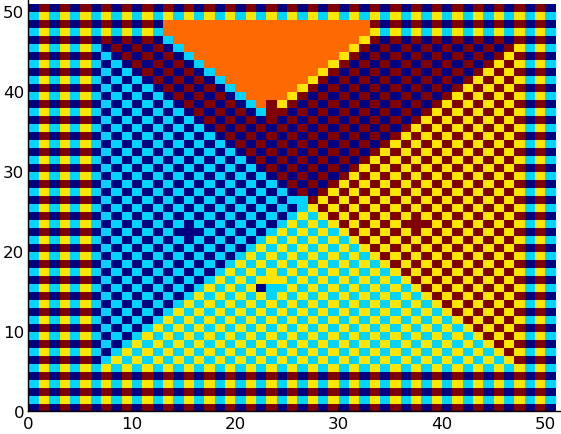}
	}
	\caption{(a) The final pattern with point perturbations in all four patches. Figure (b) shows the sites visited by the additional walker put in the top patch - the sites in orange are visited four times, while those in blue and yellow lines bounding the orange region are visited twice. Colour code: dark blue - $\rightarrow$, light blue - $\uparrow$, yellow - $\leftarrow$, red - $\downarrow$}
	\label{4type1-2}
\end{figure}

\begin{figure}[h]
	\centering
	\subfigure[]{
		\includegraphics[width=0.46\linewidth,height=0.44\linewidth]{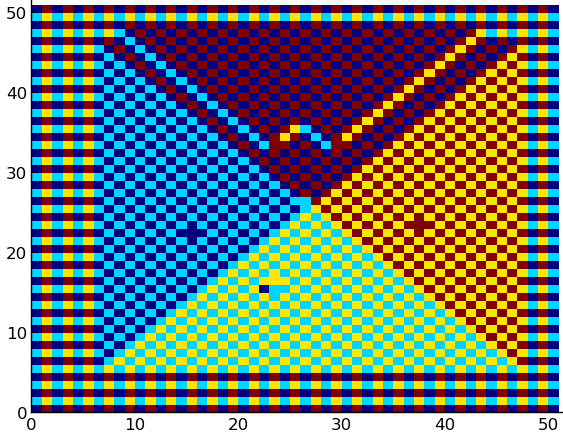}
	}
	\subfigure[]{
		\includegraphics[width=0.46\linewidth,height=0.44\linewidth]{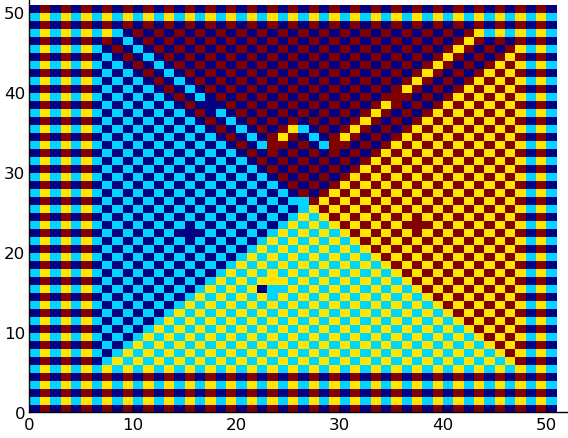}
	}
	\caption{The final pattern formed by type (i) point perturbations: (a) Two intersecting t-cones (b) Perturbed unit cells on t-cones have no effect.  Colour code: dark blue - $\rightarrow$, light blue - $\uparrow$, yellow - $\leftarrow$, red - $\downarrow$}
	\label{4type1-4}
\end{figure}

We start with the visit function for the unperturbed pattern as the initial guess: $u_0(x,y) = 40 - 2\vert x\vert$ in the top and bottom patches and $40 - 2\vert y \vert$ in the right and left patches. $\mathcal{L}~ u_0$ evaluated at all sites is shown in fig \ref{4type1-fl} (b). There is a hill of height $1$ at the site above the perturbed site and a hill of hole $1$ at the site below the perturbed site. Putting a walker at the hill and letting it walk to the hole gives the final configuration in fig \ref{4type1-2} (a), and changes $u$ as given in fig \ref{4type1-2} (b). Scanning the final configuration for cycles, we find none. Thus, the effect of the perturbation is creating the `t-cone' (that is, a conical terrace) inside which sites are visited $4$ more times than they would be in the unperturbed case.\\

We call the above specific procedure - for the hill-hole pair created by the defect, putting an extra walker at the hill and letting it walk to the hole - as `firing a defect'. We use the nomenclature `t-cone' throughout the SI to denote the V-shaped formation created by this procedure in this case. The sites on the diagonal (which are here colored yellow or light blue) will be called `sites \emph{on} the t-cone'.\\

One can also place a perturbed unit cell $\left( \besm 0 & 3 \\ 1 & 3 \eesm \right)$ in each of the other three patches and use the algorithm given in the previous section to calculate the final configuration. This is also shown in fig  - there are no t-cones generated in the other patches, and the perturbations have only a local effect.\\

Now consider the effect of several point perturbations of the same type in the top patch (change several unit cells from $\left( \besm 0 & 3 \\ 1 & 2 \eesm \right)$ to $\left( \besm 0 & 3 \\ 1 & 3 \eesm \right)$ in the initial configuration). Each defect creates a single hill and a single hole as if acting independently, since $\mathcal{L}$ is a local operator. (If a hill from one defect falls on top of a hole from another, thus ending up with $\mathcal{L}~u = 0$ at that site, we can still treat it as a hill and hole being present at the same site - a walker is stopped when it falls into the `hole', and then a new walker starts walking from the `hill'.)\\

We use the following order of firing the defects to determine the rules of interaction when more than one defect is present: fire the defects in the farthest layer from the origin first, then fire the defects in the layer below it, and so on. This procedure ensures that when we fire a defect, there are no other holes in the region that the walker from this defect explores (the t-cone region), and hence no other holes that this walker can fall into. The only hole it falls into is the one which is two sites below its hill. We make the following observations:\\

(1) It is clear that if the two points create non-intersecting t-cones, they will simply act individually.

(2) If the t-cones are nested, the visit function gets an additive part of $+4$ every time one crosses from the outside to a t-cone to inside it.

(3) If the t-cones due to two defects intersect (fig \ref{4type1-4} (a)), they merge to create a single terrace-like structure.

(4) However, if one of the points is on the boundary of a t-cone created by another, it does not create a t-cone of its own, see fig \ref{4type1-4} (b).\\

Note that these rules apply for arbitrarily big patches and arbitrary arrangements of defect unit cells. This is because of the the fact that $\mathcal{L}$ is a local operator, and the fact that even an arbitrarily long t-cone has the same structure as a small one - the region inside the t-cone that is tiled by the same unit cell as the patch, and one-unit-cell thick boundaries which continue at a $45^{\circ}$ angle until they hit the edge of the patch. This argument enables us to deduce the behaviour of larger patterns from the study of patterns formed on small lattices with a few defect sites.\\

Calculating $\mathcal{L}~u_0$ with the initial approximation $u_0$ being, as before, the visit function for the unperturbed patch, we find that the $\left( \besm 0 & 3 \\ 0 & 2 \eesm \right)$ unit cells act identically to the $\left( \besm 0 & 3 \\ 1 & 3 \eesm \right)$ ones, both producing hill-hole pairs that, when fired, generate t-cones in the patch, except the t-cones have vertices at sites where $x$ and $y$ are both odd, in one case, and at sites where $x$ and $y$ are both even, in the other. It turns out that the t-cones of the second kind of firing (due to the perturbation $\left( \besm 0 & 3 \\ 0 & 2 \eesm \right)$) interact with the t-cones of the first kind (due to $\left( \besm 0 & 3 \\ 1 & 3 \eesm \right)$) without distinguishing which kind of unit cell created the defect line.\\


We can give a heuristic explanation for the rule (4). The defect lines which form the two arms of a t-cone have the same arrow structure, locally, as the patches on either side of the current patch. That is, the defect line on the right side of a t-cone is made of the same unit cells which tile the patch to the right. We know that $\mathcal{L}$ is a local operator, and that in both these patches the perturbation has only a local effect and does not generate t-cones. Thus we can expect that a perturbation on such a defect line would also have only a local effect, and not generate t-cones either. As shown earlier, this expectation is verified by explicit studies of small systems.\\

Denote the unperturbed visit function by $V_0(x,y)$, and the visit function for a particular noise realisation by $V(x,y)$. We define a height function $h(x,y)$ in the top patch by $4 h(x,y) = V(x,y) - V_0(x,y) = V(x,y) -  (D - 2 y)$. The value of $h$ at a point $(x,y)$ which does not lie on a defect line is given by the number of terraces crossed in order to get to the region from the origin (or from the boundary of the patch). If $(x,y)$ lies {\emph{on}} a defect line, $h(x,y)$ is $h'-\frac{1}{2}$, where $h'$ is the value of $h$ for the region inside the defect line. Once we draw the defect lines according to the above rules, we can determine $h(x,y)$ for all points inside the patch, and this determines $V(x,y)$.\\

Note that the periodicity of the patches and the equivalence proved above allow us to use calculations of the final state on small-sized patches, as was done in the figures above, to generalise to large patches and arbitrary levels of noise (as long as the background stays recurrent).

\section{Type II backgrounds}

In this section, we study the patterns formed on the background:
\beqa
P\left(\left( \besm 0 & 3 \\ 1 & 2 \eesm \right)\right) = (1-p)^2 &,& P\left( \left( \besm 2 & 3 \\ 1 & 2 \eesm \right) \right) = p(1-p), \nonumber \\ 
P\left(\left( \besm 0 & 3 \\ 3 & 2 \eesm \right)\right) = p(1-p) &,& P\left( \left( \besm 2 & 3 \\ 3 & 3 \eesm \right) \right) = p^2.
\label{eq:type2}
\eeqa
On this background, defect lines are created in three patches (see Fig. \ref{4type2-wpath} (a)), and our studies show that for large $p$ the background is transient. A transition from recurrence to transience happens at $p_c \approx 0.385$.\\

\begin{figure}[h]
	\centering
	\subfigure[]{
		\includegraphics[width=0.46\linewidth,height=0.44\linewidth]{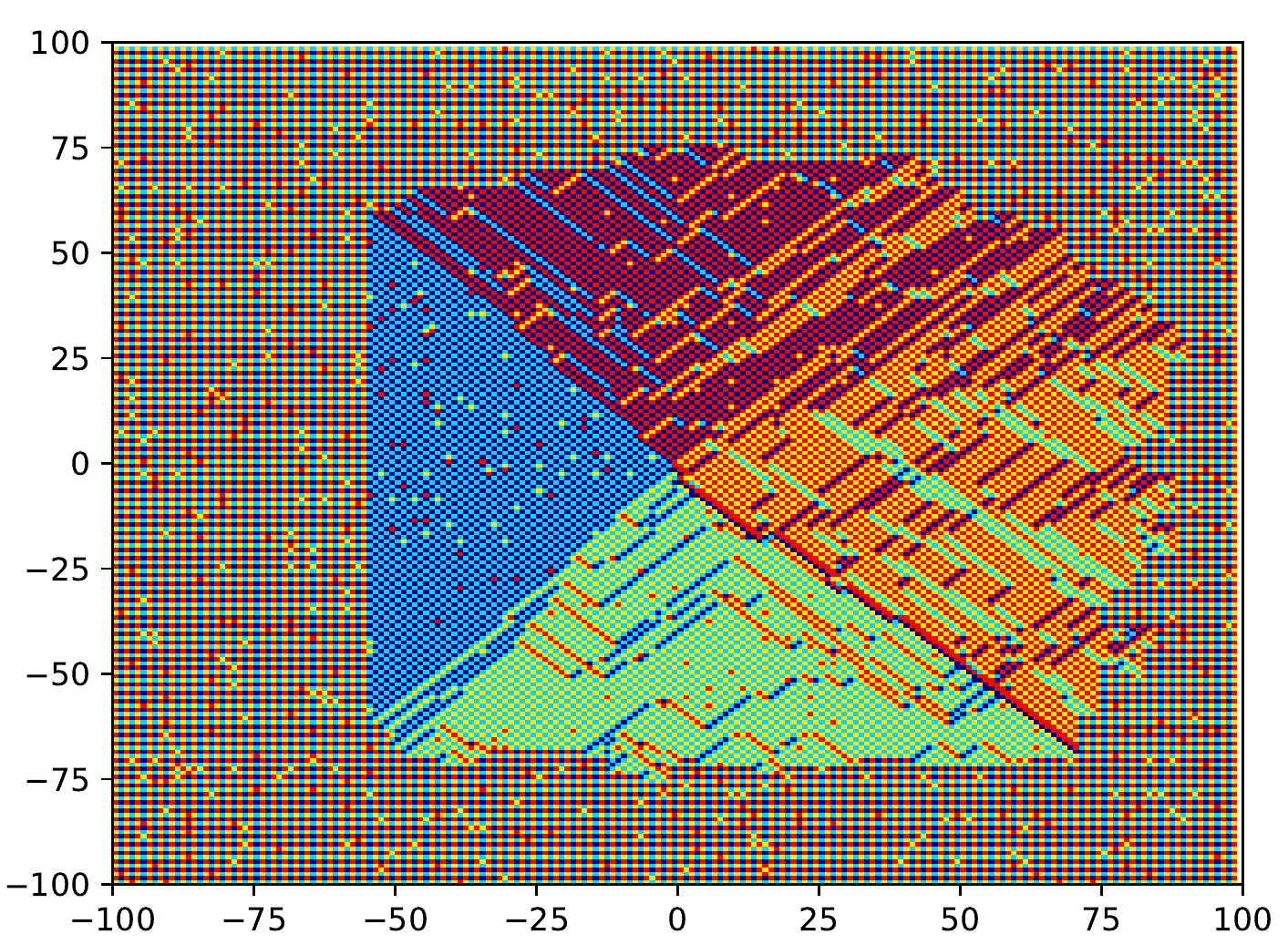}
	}
	\subfigure[]{
		\includegraphics[width=0.46\linewidth,height=0.44\linewidth]{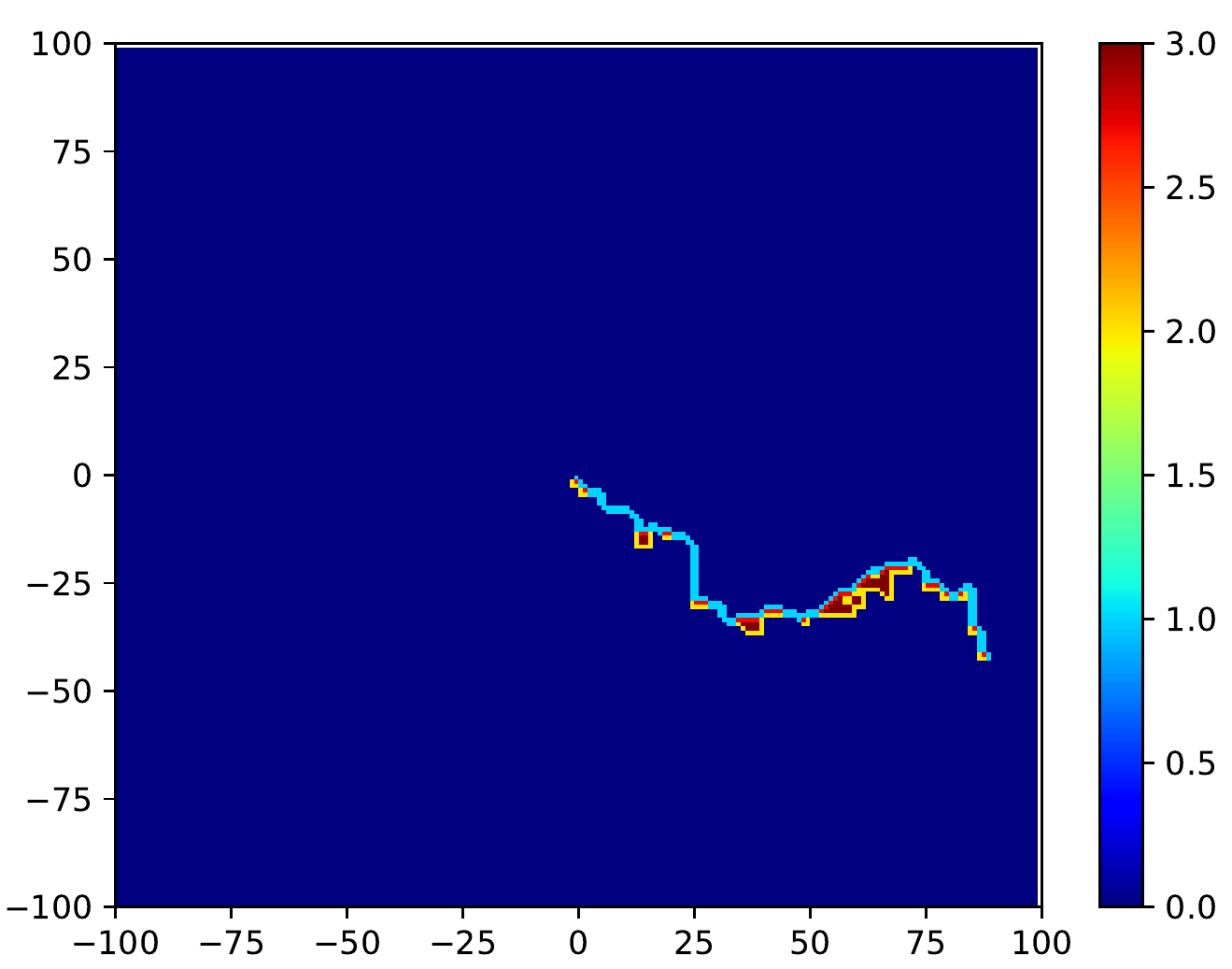}
	}
	\caption{(a) The final pattern formed by a walker on a type 2 background with $p=0.05$.  Colour code: dark blue - $\rightarrow$, light blue - $\uparrow$, yellow - $\leftarrow$, red - $\downarrow$ (b) For $p=0.45$, the sites visited by the walker, coloured by the number of visits to each site, before the walker reaches the edge of the lattice. Note that most sites are not visited, a sign of the transient nature of the walk.}
	\label{type2-wpath}
\end{figure}

Fig. \ref{type2-wpath} shows the path of the walker for two values of $p$, below and above the transition.

\subsection{Rules for drawing defect lines}

There are three kinds of defects on the type II background, `A': $\left( \besm 2 & 3 \\ 1 & 2 \eesm \right)$, `B': $\left( \besm 0 & 3 \\ 3 & 2 \eesm \right)$, and `C': $\left( \besm 2 & 3 \\ 3 & 2 \eesm \right)$. The unperturbed ($p=0$) background is the one tiled by $\left( \besm 0 & 3 \\ 1 & 2 \eesm \right)$, the same as that for case (i).\\

\begin{figure}[h]
	\centering
	\subfigure[]{
		\includegraphics[width=0.46\linewidth,height=0.44\linewidth]{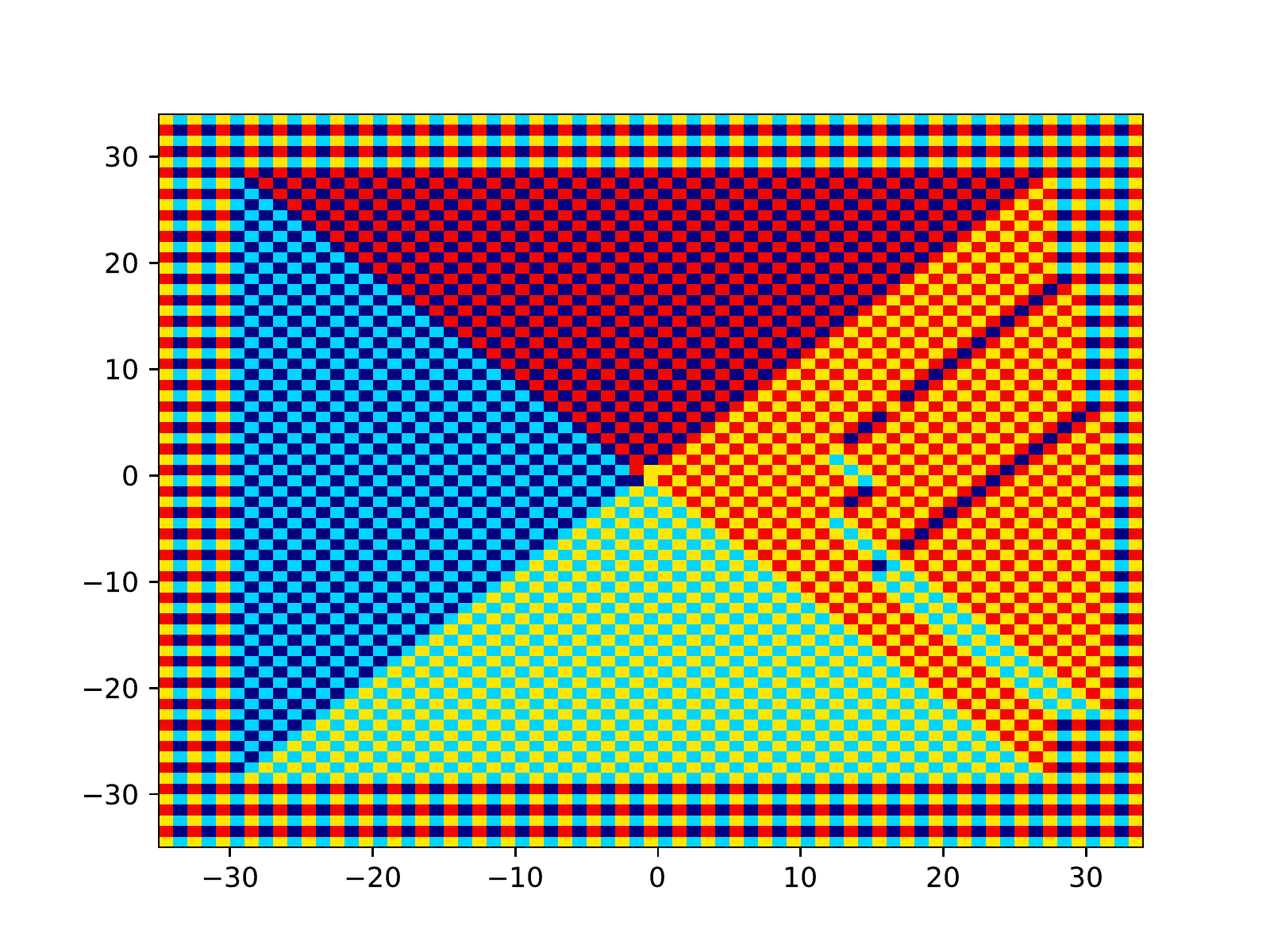}
	}
	\subfigure[]{
		\includegraphics[width=0.46\linewidth,height=0.44\linewidth]{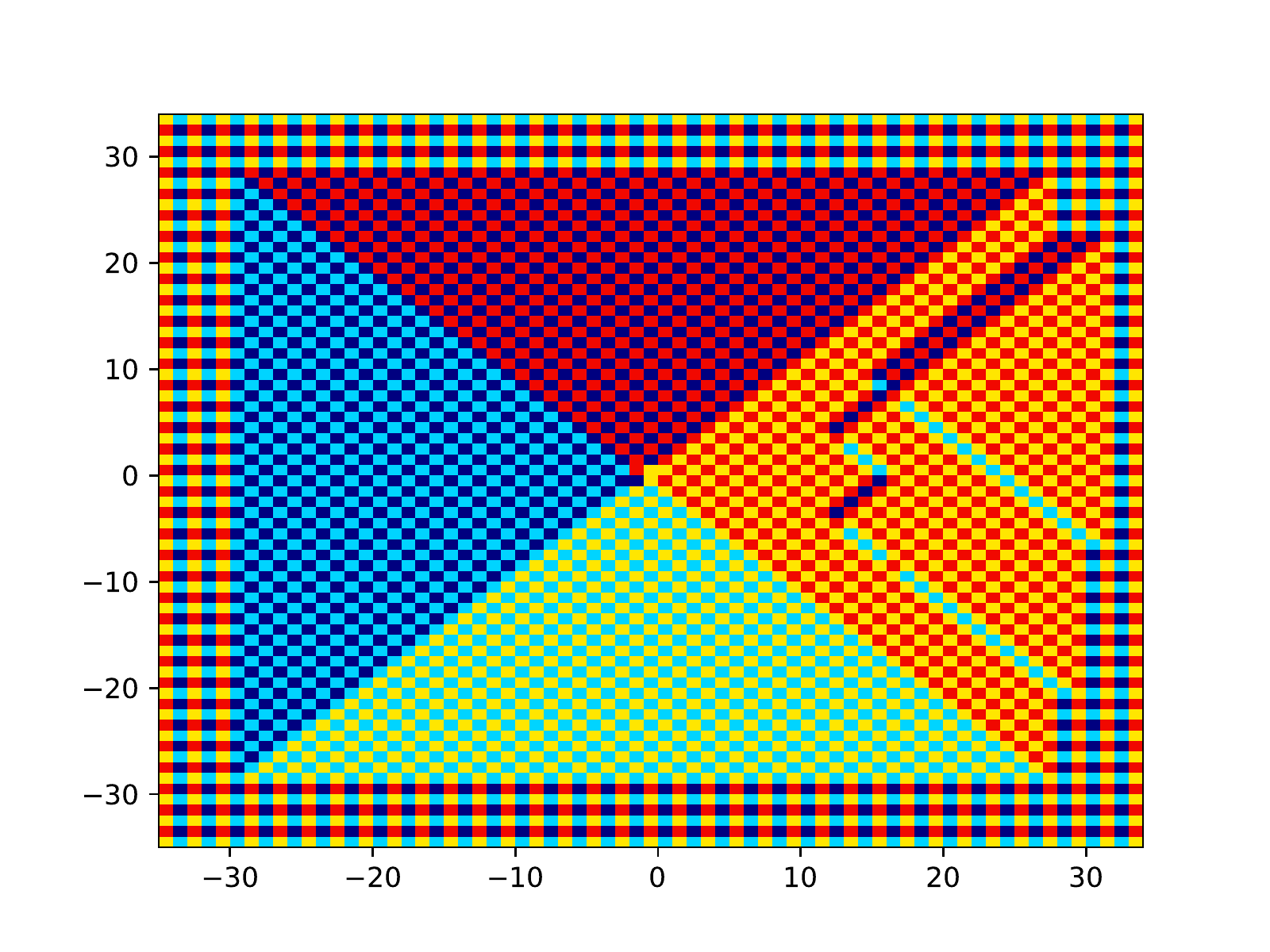}
	}
	\caption{(a) Effect of multiple point perturbations of the kind `A' (b) The effect of multiple point perturbations of the kind `B'. Colour code: dark blue - $\rightarrow$, light blue - $\uparrow$, yellow - $\leftarrow$, red - $\downarrow$}
	\label{4type2-1}
\end{figure}

Among the three types of defects, the third one is a combination of the previous two perturbations. We shall concentrate on the effect of the first two. Now let us consider the effect of a point perturbations. From Fig. \ref{4type2-1}, it can be seen that a single defect of either type `A' or `B' creates a t-cone as for the previous pattern, but the effect of multiple defect sites is not the same. The difference lies in the case where a defect lies on an existing defect line: in this case, based on the type of defect, it might create a new t-cone centered on its unit cell, and push back the existing defect line. Due to this push-back, and the anisotropic property that defects on the top defect line and the bottom defect line have different effects, an exact mapping to a solvable model is not possible.\\

However, the height function can be defined similarly as for the previous case, such that it changes by $4$ on crossing into a terraced region, as previously. Thus a height field $h(x,y)$ can be defined, and calculated by drawing the terraces based on the positions on the defects. This height field then gives the final visit function, $V(x,y) = V^0(x,y)+h(x,y)$, for $p<p_c$.

